\documentclass[sigconf]{acmart}

\AtBeginDocument{%
  }

\usepackage{graphicx}
\usepackage{multirow}
\copyrightyear{2024} 
\acmYear{2024} 
\setcopyright{acmlicensed}\acmConference[CVMP '24]{European Conference on Visual Media Production}{November 18--19, 2024}{London, United Kingdom}
\acmBooktitle{European Conference on Visual Media Production (CVMP '24), November 18--19, 2024, London, United Kingdom}
\acmDOI{10.1145/3697294.3697306}
\acmISBN{979-8-4007-1281-4/24/11}

\acmSubmissionID{12}

\begin{document}

%%
%% The "title" command has an optional parameter,
%% allowing the author to define a "short title" to be used in page headers.
\title{PDFed: Privacy-Preserving and Decentralized Asynchronous Federated Learning for Diffusion Models}

%%
%% The "author" command and its associated commands are used to define
%% the authors and their affiliations.
%% Of note is the shared affiliation of the first two authors, and the
%% "authornote" and "authornotemark" commands
%% used to denote shared contribution to the research.
\author{Kar Balan}
\email{k.balan@surrey.ac.uk}
\orcid{0009-0004-1296-9254}
\affiliation{
  \institution{University of Surrey}
  \city{Guildford}
  \country{United Kingdom}
}

\author{Andrew Gilbert}
\email{a.gilbert@surrey.ac.uk}
\orcid{0000-0003-3898-0596}
\affiliation{%
  \institution{University of Surrey}
  \streetaddress{Stag Hill, University Campus}
  \city{Guildford}
  \country{United Kingdom}
  \postcode{GU2 7XH}
}

\author{John Collomosse}
\email{collomos@adobe.com}
\orcid{0000-0003-3580-4685}
\affiliation{%
  \institution{Adobe Research}
  \city{San Jose}
  \country{USA}}

%%
%% By default, the full list of authors will be used in the page
%% headers. Often, this list is too long, and will overlap
%% other information printed in the page headers. This command allows
%% the author to define a more concise list
%% of authors' names for this purpose.
\renewcommand{\shortauthors}{Balan et al.}

%%
%% The abstract is a short summary of the work to be presented in the
%% article.
\begin{abstract}
We present PDFed, a decentralized, aggregator-free, and asynchronous federated learning protocol for training image diffusion models using a public blockchain. In general, diffusion models are prone to memorization of training data, raising privacy and ethical concerns (e.g., regurgitation of private training data in generated images). Federated learning (FL) offers a partial solution via collaborative model training across distributed nodes that safeguard local data privacy. PDFed proposes a novel sample-based score that measures the novelty and quality of generated samples, incorporating these into a blockchain-based federated learning protocol that we show reduces private data memorization in the collaboratively trained model. In addition, PDFed enables asynchronous collaboration among participants with varying hardware capabilities, facilitating broader participation. The protocol records the provenance of AI models, improving transparency and auditability, while also considering automated incentive and reward mechanisms for participants. PDFed aims to empower artists and creators by protecting the privacy of creative works and enabling decentralized, peer-to-peer collaboration. The protocol positively impacts the creative economy by opening up novel revenue streams and fostering innovative ways for artists to benefit from their contributions to the AI space.

\end{abstract}


%%
%% Keywords. The author(s) should pick words that accurately describe
%% the work being presented. Separate the keywords with commas.
%\keywords{Do, Not, Us, This, Code, Put, the, Correct, Terms, for, Your, Paper}
%% A "teaser" image appears between the author and affiliation
%% information and the body of the document, and typically spans the
%% page.
\begin{teaserfigure}
    \includegraphics[width=\textwidth]{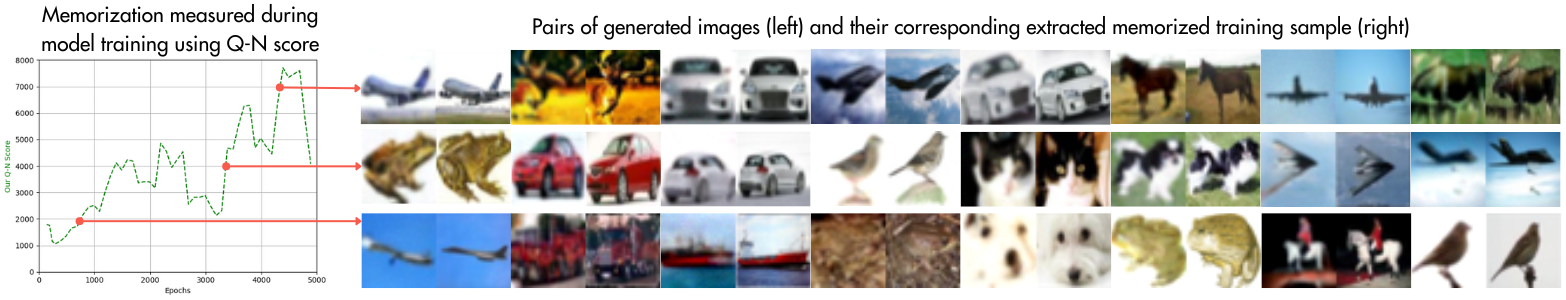}
    \caption{Data memorization presents a privacy challenge for the federated training of diffusion models, where private data can be regurgitated as training progresses. PDFed proposes a novel measure and mitigation of this behavior through an aggregator-free distributed training protocol using distributed ledger technology (DLT), or `blockchain' that we later show reduces data memorization.}
    \label{teaser}
\end{teaserfigure}

%%
%% This command processes the author and affiliation and title
%% information and builds the first part of the formatted document.
\maketitle

\section{Introduction}
Diffusion models are a powerful class of generative AI (GenAI) algorithms transforming creative practice by democratizing the ability to create realistic and readily customizable visual content. Diffusion models require training on vast datasets to achieve state-of-the-art performance. However, collecting and using such data poses significant challenges, including privacy, copyright, and creator consent. In particular, diffusion models are prone to  {\em training data memorization} and often regurgitate training data within generated content. This poses data privacy risks, particularly for copyrighted creative works or identifiable personal images. As a result, there is a growing demand for privacy-preserving methods that enable ethical diffusion model training while respecting data privacy. 

Federated learning (FL) is a machine learning approach that enables model training across decentralized edge devices or servers, each holding local data samples without exchanging or exposing them. FL emerged as a solution to the pressing demand for individuals and organizations to maintain control over their data while still participating in the collective training of AI models. Integrating diffusion models with FL allows for collaborative training across distributed datasets while safeguarding data privacy and security, unlocking new opportunities for models to access and learn from diverse datasets that would otherwise be inaccessible. Existing FL \cite{behera2022federated, blockflow, bcafl} approaches typically require a trusted, centralized node to orchestrate the training process, including aggregating model submissions and disseminating updates to participating nodes. This centralized node can become a single point of failure and a privacy risk if compromised, and its operations may lack transparency. Constant communication between the central and participating nodes also introduces a significant communication overhead. This overhead, coupled with a requirement for new participants to wait for the aggregation result until the current training round concludes, introduces latency.

\emph{PDFed} addresses these issues by leveraging federated learning principles while introducing a decentralized and asynchronous framework to enhance privacy preservation, model performance, and training efficiency. Reliance on a centralized server is eschewed in favor of orchestrating the training process via distributed ledger technology (DLT), namely the Ethereum public `blockchain'. To address the phenomenon of training data memorization in diffusion models, we propose and integrate a parametric sample-based score into our training framework, which measures sample novelty, fidelity, and quality. PDFed makes the following two technical contributions:

1. \textbf{Asynchronous, aggregator-free and decentralized federated learning protocol.} PDFed is the first federated learning framework for training diffusion models that does not require a central coordinator or aggregator. Instead, we take advantage of the decentralized, leaderless nature of public blockchains, allowing participants to operate asynchronously and independently according to a local training strategy. We also permit validation-only participants, enabling agents who hold data but don't have the hardware resources necessary for active model training to contribute by voting for the best scoring of the submitted models, facilitating broader participation.

2. \textbf{Quality-novelty (Q-N) objective.} To guide model training, a Q-N score is proposed. Generated samples undergo a visual fingerprinting step, followed by a KNN search to obtain a shortlist of samples close to the training data, implying local memorization. The pairs of generated images and training samples are then passed to a second model, which outputs a content similarity score for each pair. The proposed metric is based on this content similarity score and the FID score of all generated samples. The proposed metric acts as an objective function within the training framework, aiding in selecting models for further training, favoring those with lower memorization levels, and successfully driving a trend towards reduced memorization. Additionally, the score is leveraged at the data loader level, ensuring that previously identified memorized training images are excluded from further use in model training. This method effectively acts as a decentralized, privacy-preserving training data deduplication system in the distributed training scenario.

\section{Related Work}

\textbf{Diffusion models} have enabled recent advancements in Generative AI, producing realistic samples across various modalities such as images and video. Despite requiring datasets comprising millions of samples for training, they have been observed to exhibit training data memorization, particularly in content and style, attributed to duplicated data within the training dataset. Carlini et al. \cite{carlini2023extracting} and Somepalli et al. \cite{somepalli2022diffusion} confirm memorization in diffusion models across datasets of varying scales and conditioning types and demonstrate that training data deduplication is effective in reducing training data memorization. In a distributed, privacy-preserving training scenario, participants cannot know whether they hold duplicate data as other participants without compromising their data privacy. There's a research gap in privacy-preserving, distributed image data deduplication. Stein et al. \cite{stein2023exposing} and Yoon et al. \cite{yoon2023diffusion} further explore the conditions that contribute to memorization and identify several influential factors such as random, uninformative, or highly specific labels, model capacity, dataset size and complexity, distribution of rare-common images and concepts, and type of conditioning. EKILA \cite{balan2023ekila} introduces a framework for identifying the training samples most influential for a specific generated image, enabling the authors of those training images to be rewarded for their contributions.

%images (DALL-E 3, Imagen 2, and Stable Diffusion 2) and video (stable video diffusion, Lumiere, FateZero)

\textbf{Memorization metrics.} Several metrics have been proposed to assess the novelty, diversity, and fidelity of diffusion model-generated samples. AuthPct \cite{authpct} deems each generated sample either as authentic or inauthentic based on whether the distance to the nearest point in the training set is less than the distance between that training sample and its nearest neighbor in the training set. The $C_T$ score \cite{ctscore} summarizes how often training samples are closer to generated samples than to samples in the test set through a Mann-Whitney hypothesis test\cite{mann}. Jiralerspong et al. \cite{jiralerspong2023feature} propose the FLD score, a parametric sample-based score that relies on density estimation in feature space to compute the perceptual likelihood of generated samples. \cite{stein2023exposing} finds that AuthPct \cite{authpct}, $C_T$  \cite{ctscore}, and FLD \cite{jiralerspong2023feature} scores are only sensitive to memorization in an ideal scenario where all samples become increasingly memorized and that AuthPct and $C_T$ scores detect mode shrinking and image fidelity more than memorization. 
%Further, \cite{stein2023exposing} argues that replacing Inception-V3\cite{inceptionv3} as the feature extractor with DINOv2 ViT-L/14\cite{dinov2} for calculating the FID score \cite{fid} solves the discrepancy with human evaluators, a measure we adopt in our experiments.

\textbf {Distributed Ledger Technology} (the most well known implementation is 'blockchain') is a decentralized system that facilitates the secure and transparent recording of transactions across multiple nodes or participants. Unlike traditional centralized systems, DLT does not rely on a single point of control or trust. Instead, it enables consensus mechanisms that allow all participants to agree on the validity of transactions without the need for intermediaries or central authorities. While initially popularized by cryptocurrencies such as Bitcoin, the applications of DLT have evolved beyond finance, with many innovative applications for social good. DLT has been used in digital preservation, where it ensures the integrity and immutability of data over time\cite{archangel}, and in AI training, to determine author consent for specific data to be used\cite{decorait}. Smart contracts are self-executing programs automatically enforced and executed on a blockchain network without intermediaries. The event log is a ledger of signals or exceptions emitted from smart contract code.

\textbf{Federated Learning}  addresses data privacy by collaboratively training models without pooling or exchanging the training data. \cite{feddisc} and \cite{noniid} propose one-shot FL frameworks in which workers train diffusion models using their privately held data, such that data can subsequently be generated conforming to the global class distribution to alleviate the non-IID data problem. 
 %In \cite{oneshot}, a small diffusion model is subsequently trained with differential privacy using the previously generated data and is found not to exhibit signs of training data memorization, attributed to the small number of parameters of the chosen model architecture and the differential privacy training technique.
 \cite{sensitivevis} explores several FL scenarios for training diffusion models, investigating the effects of varying the data heterogeneity across clients, the number of clients, and the number of training rounds and epochs; they obtain comparable results to centralized training and find that distinct data distributions lead to local models being biased towards local data, which can challenge model aggregation and affect global model performance in early FL rounds. %however, these local models may act as personalized models in later rounds. 
The inherent technical capabilities of DLT make it a natural choice for integrating with FL frameworks. \cite{bcafl} proposes building a specialized blockchain with a bespoke consensus mechanism for a federated learning task. %\cite{behera2022federated} implements a federated learning framework on a consortium blockchain, where one node acts as a federated aggregation server, and others act as federated clients. Smart contracts are used to maintain records of contributions, and a unique scalar quantity is calculated to quantify each participant's contribution. 
\cite{competitive} abolishes the concept of rounds, allowing workers to operate independently without the need for synchronization or a central aggregator. However, nodes' participation is time-limited. \cite{baffle} proposes an aggregator-free FL system, storing the global model as serialized chunks within a smart contract. Workers place bids to update specific chunks, and the highest-scoring one pushes the update to the smart contract. In Blockflow\cite{blockflow}, all client nodes report evaluation scores for all submitted models, which the smart contract uses to calculate the weights for model aggregation conducted locally by all clients. Blockflow enforces strict deadlines during rounds, removing client nodes that are slower to submit. % thus missing out on potential contributions, a disadvantage of synchronous FL. \cite{audit} introduces a data-model provenance registry within the smart contract, which maintains an auditable provenance chain for the model submissions to improve AI fairness and accountability. 
None of these solutions explore DLT for training diffusion models or seek to mitigate diffusion memorization specifically.
% \cite{short} proposes a system in which the smart contract is responsible for verifying contributions, calculating the global model, and recording users' performance to support the incentive scheme. The verification function would be run using an off-chain service. However, the proposed system has not been implemented in practice.
\section{Measuring and Mitigating Memorization}
\label{sec:metric}
PDFed is a privacy-preserving, decentralized, federated learning framework for training diffusion models while successfully reducing private data memorization and building an auditable model provenance chain, significantly contributing to transparently building ethical AI models. It is asynchronous and aggregator-free, each node employing a local training strategy. We now describe PDFed, first focusing on the proposed methodology for measuring memorization in the trained model, which is used within the FL process (Sec.~\ref{sec:fed}) to govern model validation and selection.

We propose a method to identify memorized training samples and a parametric sample-based score that measures the novelty, fidelity, and quality of images generated by GenAI models. The score may be implemented as a loss function in GenAI training, allowing monitoring and measuring the memorization rate and optimizing training to minimize this behavior. The score is based on a content similarity score between training data and generated samples and the widely utilized FID score. Obtaining the content similarity score is a 3-step process: 1) establishing a distance threshold based on statistical properties of the training data to identify generated samples unusually close to the training data for further analysis, 2) partial matching based on image fingerprints extracted from generated and training data and 3) pairwise verification and scoring of retrievals from the previous step.

\subsection{Model Architectures}

\textbf{Image fingerprinting.} Similar to \cite{balan2023ekila}, we adapt the visual fingerprinting approach introduced by \cite{Black_2021_CVPR} to generate compact, 256-dimensional embeddings of the generated and training images, enabling retrieval of visually similar pairs at scale. The fingerprinting model is trained using a contrastive learning approach \cite{chen2020simple} to be discriminative of image content while robust to image degradations and manipulations.

Given an image $x_i$, we denote its embedding obtained from a ResNet-50 model as $\phi_i = E(x_i) \in \mathbb{R}^{256}$, with $\hat{\phi}i$ representing an augmented version of the same image. The training objective is:
\begin{equation}
\mathcal{L}_{C} = - \sum_{i\in \mathcal{B}}   \log \left( \frac{d\left( \phi_i, \hat{\phi}_i \right)}{ d\left( \phi_i, \hat{\phi}_i \right) + \sum_{j \neq i \in \mathcal{B}} d\left( \phi_i, \phi_j \right)} \right),
\end{equation}
where $d(a, b):= \exp \left(\frac{1}{\lambda}  \frac{ {a}^\intercal { b}}{\Vert {a} \Vert_2 \Vert {b} \Vert_2} \right)$ measures the similarity between embeddings, and $\mathcal{B}$ represents a randomly sampled mini-batch during training \cite{balan2023ekila}.

\textbf{Image match verification.} A shortlist of the top-K candidate image pair matches is obtained using the previously extracted image fingerprints. The generated query image and each candidate match retrieved from the training dataset are verified through an additional pairwise comparison. The spatial feature maps derived from the fingerprinter model are used to compare the image pairs as presented in \cite{balan2023ekila}. During training for the match verification model, the backbone feature extractor model is frozen.

Let $F_q\in \mathbb{R}^{H\times W \times D}$ be the feature map for a query image $x_q$ and let $\{F_i\}_{i=1}^k$ be the $k$ corresponding retrieval feature maps.
Each feature map is processed with a $1\times1$ convolution to reduce the dimensionality to $\frac{D}{4}$ and then numerous pooled descriptors from a set of 2D feature map windows $\mathcal{W} \subset [1, H] \times [1, W]$ are extracted, similar to R-MAC \cite{tolias2015particular}.
Let $f^q_{w}\in \mathbb{R}^\frac{D}{4}$ denote the GeM-pooled \cite{tolias2015particular} and unit-normalized feature vector for a window $w\in \mathcal{W}$ and feature map $F_q$. The window-pooled feature vectors are collected as follows:
\begin{equation}
    \hat{F}_q = [f^q_{w_1}, \ldots, f^q_{w_{|\mathcal{W}|}}] \in \mathbb{R}^{|\mathcal{W}| \times \frac{D}{4}},
\end{equation}
with $|\mathcal{W}| = 55$ windows in practice and $w_i \in \mathcal{W}$.
The correlation matrix between the features of the query and candidate matches is computed as:
\begin{equation}
    C_{qi} = \hat{F}_q \hat{F}^T_i \in \mathbb{R}^{|\mathcal{W}|\times |\mathcal{W}|}.
\end{equation}
This is then flattened and fed to a 3-layer $\operatorname{MLP}$, which outputs the similarity score between the query and retrieval images.
To make the model symmetric w.r.t. its inputs, the similarity score between images $x_q$ and $x_i$ is defined as follows, where $\sigma$ is a sigmoid activation.
\begin{equation}
    \operatorname{score}(x_q, x_i) = \sigma \big( \operatorname{MLP}(C_{qi}) + \operatorname{MLP}(C_{iq}) \big),
    \label{eq:apportion}
\end{equation}

Data augmentation, such as color jittering and random cropping, as well as minor, benign modifications, manipulations, and degradation of image content due to noise, format change, compression, and resolution change are applied to enhance robustness against benign image alterations \cite{hendrycks2019robustness}. This is to model artifacts commonly present in generated images during the initial and later phases of training GenAI models, such as blurred and distorted images. Despite the artifacts, this approach allows us to identify images with an unusually similar style or content as training data. Positive training pairs are created using augmented samples, while challenging negative pairs are generated via hard negative mining. Global average-pooled feature maps of query and queued examples are compared via cosine similarity for the sampling of negatives. The model is trained using binary cross-entropy loss on pairs of true and false matches.

\subsection{Identifying memorized samples}

\textbf{Intra-class threshold for retrieval distance.} Carlini et al. \cite{carlini2023extracting} propose a method to identify memorized samples strictly based on L2 distance thresholding. The memorized image extraction method considers an image as memorized if its L2 distance to the nearest neighbor in the training set is significantly lower than that of all other training images. 
%This distance is calculated based on the extracted image's proximity to its nearest training image relative to the training image's nearest neighbors.

We draw inspiration from this approach and compute intra-class distance thresholds rather than individual image thresholds. We extract image fingerprints from all samples in the training set and calculate the L2 distance between each image and its closest intra-class neighbor. We compute the mean and standard deviation of this distance across each class to determine the threshold \( T_{\text{L2}_{\text{class}}} \). Any generated image whose L2 distance to a training sample falls below this threshold is deemed unusually close to the training data, prompting further analysis. 

\begin{equation}
T_{\text{L2}_{\text{class}}} = \text{mean} - ( 0.5 \times \text{stdev})
\end{equation}

\textbf{KNN search.} During this step, a KNN search is performed between each generated sample and local training samples to determine the pairs that fall under the intra-class threshold. The list of pairs is compiled and passed to the next step.

Unlike Carlini et al. \cite{carlini2023extracting}, our approach incorporates L2 search but does not solely rely on it. Instead, it leverages a bespoke feature extractor and integrates a secondary pairwise verification stage, as outlined below.

\textbf{Image match verification.} During this step, each pair consisting of a generated sample and a retrieved, unusually similar training sample is passed to the image match verification model, resulting in a similarity $\operatorname{score}(x_q, x_i) \in [0,1]$. We empirically set a 0.8 threshold for an image match confirmation. There are often duplicates within training data, and some samples are generated multiple times. Due to this, each generated sample and training sample may be counted as memorized only once. This step concludes with a final count of memorized samples.

\subsection{Metric design}

Our metric balances sample quality and fidelity against novelty to robustly evaluate diffusion models.

\textbf{Novelty.} We define novelty as the degree to which generated samples differ from the training samples. Memorized samples are simply reproductions of training samples. The three values that determine the novelty component of our proposed metric are calculated using similarity scores output by the match verification model as such:

\textbf{1) $V_A$: The mean similarity score of all checked pairs.} This value encompasses the mean similarity score of all pairs passed to the image match verification model, whether confirmed or unconfirmed matches. We consider that a novelty score shouldn't be influenced solely by those samples that are unquestionably memorized. Instead, all pairs passed to the verification model exhibit unusual similarity to training samples and thus should be accounted for when designing a robust and multidimensional novelty score. Although unconfirmed, unusually similar samples to training data may indicate early signs of memorization.

\textbf{2) $V_C$: The mean similarity score of all confirmed matches.} This introduces a measure of the similarity between memorized and training samples, thereby enhancing accuracy compared to a binary true/false measure.

\textbf{3)$R_C$: The ratio of confirmed memorized samples to the total number of generated samples.}

\textbf{Quality and fidelity.} To account for the quality and fidelity of generated samples, our metric incorporates the FID score calculated between the generated samples and each node's local test samples. The feature extractor utilized is DINOv2 ViT-L/14 \cite{dinov2}, replacing the traditional Inception-V3, as Stein et al. \cite{stein2023exposing} found this replacement solves the discrepancy with human evaluators. Consequently, the FID value is notably higher than expected due to this change.

Our proposed quality-novelty score is computed as follows:
\begin{equation}
\label{ours}
\text{Q-N score} =\frac{{\text{FID} + (V_C \times V_A \times R_C \times 1000)}}{2}
\end{equation}
Our Q-N score is customizable - the user may adjust parameters such as the intra-class distance and match confirmation threshold. Unlike concurrent metrics such as FLD, which necessitates a minimum of 10,000 generated samples for an accurately reported score, the Q-N score can be computed irrespective of the train, test, or generated image set size. Our score also allows for the extraction of memorized sample pairs, unlike other metrics, which rely on the statistical properties of the data to identify and measure memorization rather than conducting pairwise comparisons. The computational overhead of calculating our score is minimal. Fingerprinting, searching for matches, and verifying an image pair typically take an average of 41 ms per image. The fingerprinter model requires a minimum of 1 GB of GPU memory, while the verifier model requires 2.6 GB of GPU memory.

\section{Federated Learning Framework}
\label{sec:fed}

\begin{figure*}[h]
    \centering
    \includegraphics[width=\textwidth]{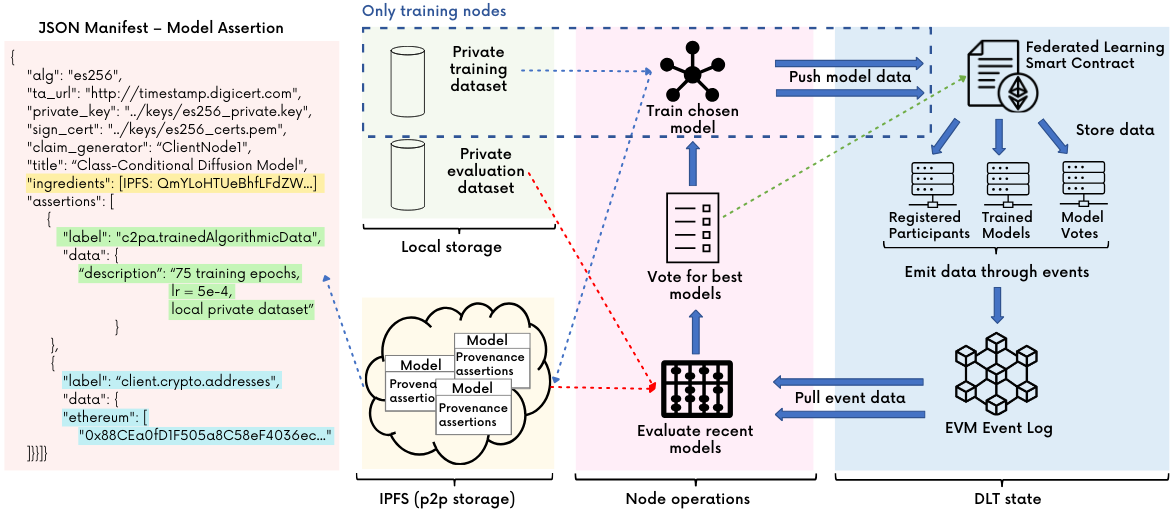}
    \caption{System Architecture: client nodes, holding private training/validation data, interact with the federated learning task smart contract deployed on an Ethereum blockchain. The system design allows nodes to participate asynchronously, contributing to model evaluation and selection while retaining autonomy over training strategies and resource usage. The model weights are stored on peer-to-peer solution IPFS, accompanied by C2PA \cite{c2pa} manifests. An example manifest is included on the left - a client node describes a model contribution within PDFed. Highlighted in yellow - C2PA supports specifying ingredient assets; in this case, the IPFS storage CID of the model chosen for further training is included. Green - this designation enables the client node to describe the submitted model's training steps. Blue - using the crypto.addresses assertion, the client node specifies their blockchain wallet address, where payments may be sent as a reward for their contributions to model training.}
    \label{fig:system}
\end{figure*}

We introduce an asynchronous, decentralized, and aggregator-free smart contract-based federated learning protocol explicitly tailored for diffusion models, aiming to tackle prevalent issues in GenAI and traditional FL. As depicted in Figure \ref{fig:system}, a smart contract is deployed on a a public blockchain (for experimental purposes we chose the Ethereum blockchain network), orchestrating the federated learning process. The smart contract facilitates the registration of new client nodes and the submission of new models and votes. It emits events to notify all listening nodes of these submissions. Events emitted by the smart contract are transparent and immutable, providing a record of all actions and decisions made during the FL process, effectively replacing the centralized server in traditional federated learning and enhancing the system's overall resilience. Our protocol accommodates dynamic participation, enabling nodes to join or leave the training process at any point without disrupting the overall workflow.
%eliminating both the risk of a single point of failure and the communication overhead, thus enhancing the system's overall resilience.
%Client nodes continuously listen for events emitted by the smart contract on the blockchain, which notifies them of new model submissions or votes for past models. 
 %Traditional federated learning typically operates within predefined rounds to organize the training, validation, and model update dissemination cycle. However, client nodes may have different hardware configurations, including varying processing power, GPU capacity, and network bandwidth. 

Each client node holds local data samples and contributes to the training or validation of the diffusion models. They evaluate model submissions on their privately held data using our proposed Q-N metric and submit votes indicating the best-performing model updates. The other participants then consider these votes in conjunction with their evaluations when selecting models for further training. The chosen model is further trained using the client node's private training set, which remains private and does not leave their instance.

The latency incurred by on-chain operations is minimal, as most operations are read-only, resulting in near-instantaneous responses. The only essential transaction involving writing to the blockchain is the model submission, which is typically processed within seconds, depending on network congestion.

\subsection{Local strategy}

Traditional federated learning frameworks struggle to accommodate hardware heterogeneity among client nodes, leading to sub-optimal performance and resource utilization. Time-limited training rounds lead to missed opportunities for valuable data contribution and participation. In contrast, our framework functions asynchronously, allowing client nodes to participate in training and validation activities according to a local strategy tailored to their resources.

Our aggregator-free protocol decentralizes the model selection process, granting all nodes access to every model update. This protocol empowers participants to evaluate all available updates independently and determine which ones to aggregate and further train. It protects against malicious actors, as underperforming models will not be picked up for additional training. Clients can also customize their strategy by deciding which votes to consider, whether they prefer input from specific actors (training nodes or validation-only nodes) or a combination of both. Moreover, clients can adjust their training process according to their resource availability, training for as long or as little as necessary.

Another novel feature of PDFed is the support for validation-only nodes, which evaluate model updates and submit votes based on their assessment. They may have limited hardware capabilities but possess valuable data that can contribute to the validation of model updates, providing an additional layer of scrutiny, enhancing the robustness of model evaluation and selection and overall effectiveness of the protocol. This inherent flexibility facilitates adoption and broader participation, ensuring optimal performance across diverse hardware configurations and training scenarios. 

\subsection{Model Auditability and Incentives}
A key component of PDFed is the maintenance of an auditable chain of model provenance, facilitated by emerging open standard from the Coalition for Content Provenance and Authenticity \cite{c2pa}) and the corresponding C2PA tool, which allows model authors to bind provenance information to model files via cryptographically signed asset `manifests.' These manifests, alongside model files, are stored within a distributed file system (IPFS). C2PA manifests may bear information about "ingredient" assets used in the training process, such as a summary or hash of the private training data and the base model that was further trained. The base model is accompanied by its unique manifest and all its previous training assertions, which are then included in the new model's provenance manifest. For each subsequent model submission, the client node adds a new training assertion that contains authorship and training facts and the author's blockchain wallet address. This effectively builds a unique model provenance graph, which grows with each new model submission, ensuring transparency within the training process. An example manifest is pictured in Fig. \ref{fig:system}.

In conjunction with each model's C2PA manifest detailing its provenance, the smart contract enables comprehensive tracking of all model contributions. Rewards may be apportioned based on the contributions made by each client node, as detailed in the resulting model's manifest. The smart contract can automatically distribute payments to participants' blockchain wallets to incentivize and reward active participation and valuable contributions from client nodes.

%could add c2pa info into related work and mention openAI generated created image there is, but not enough space left
\begin{figure*}[ht!]
    \centering
    \includegraphics[width=\textwidth]{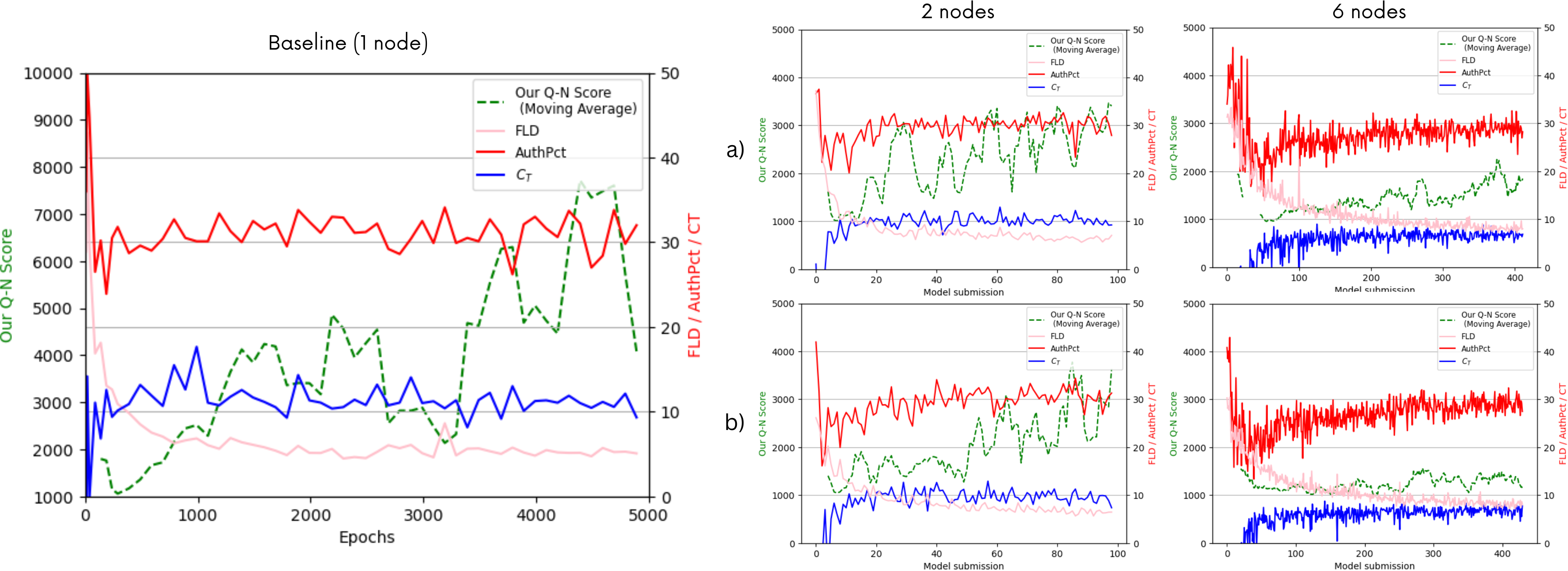} 
    \caption{\textbf{Left:} All metrics reported on the baseline experiment evaluated along the training process on the entire CIFAR-10 dataset. \textbf{Right:} a) were trained with the objective $L_{FLD+FID}$ and b) were trained with the objective $L_{Q-N}$, for experiments with 2 and 6 client nodes. The scores represented are Q-N score (green), AuthPct (red), FLD (pink), and $C_T$ (blue). Our proposed training method using $L_{Q-N}$ decreases memorization and our proposed Q-N metric is more sensitive to recognizing memorization behavior than all other metrics.}
    \label{fig:baseline}
\end{figure*}
\section{Evaluation}
\subsection{Experimental Setup}
We demonstrate PDFed as deployed through a smart contract on a public Ethereum blockchain and evaluate its performance across several experiments. The chosen task is to train class-conditional DDPM \cite{hddpm} diffusion models. For this task, we utilize the CIFAR-10 dataset \cite{cifar10}, a commonly used benchmark dataset in concurrent works \cite{carlini2023extracting, somepalli2022diffusion, jiralerspong2023feature, competitive} and widely used in many other computer vision tasks.  The dataset is divided equally into training and test sets for each client node, with training setups involving cohorts of 1 (baseline), 2, 4, 6, and 8 client nodes. An equal number of images from each class are randomly distributed among the client nodes for the private data split. The participants train each model for an equal number of epochs as each other before model submission (2 nodes: 50 epochs; 4, 6 nodes: 75 epochs; 8 nodes: 100 epochs) to ensure the models have the same degree of exposure to the data across experiments. Each node generates 1,000 images spread equally across all classes for model evaluation. All model submissions are evaluated by each client node, reporting all metrics as calculated over their privately held, local dataset split. In addition, we compute all metrics for all model submissions as evaluated on the entire CIFAR-10 dataset. 

We design three training setups to evaluate the performance of our proposed training framework in reducing memorization behavior in GenAI models: 1) $L_{FLD+FID}$, 2) $L_{Q-N}$ and 3) $L_{Q-N+dedup}$.

In the first training setup, referred to as $L_{FLD+FID}$, the loss function used by the participants when evaluating and selecting the top-performing model for further training consists of a blended FID and FLD score\cite{jiralerspong2023feature}, as given by:

\begin{equation}
\text{$L_{FLD+FID}$} =\frac{{\text{FID} + (\text{FLD} \times 100)}}{2}
\end{equation}

\begin{table*}
\centering
\normalsize
\resizebox{\textwidth}{!}{
\begin{tabular}{|l|c|c|c|c|c|c|c|c|c|c|c|c|c|c|c|}
\hline
\multirow{2}{*}{\textbf{\#}} & \multicolumn{5}{c|}{\textbf{Training run with} $\mathbf{L_{FLD+FID}}$ } & \multicolumn{5}{c|}{\textbf{Training run with} $\mathbf{L_{Q-N}}$} & \multicolumn{5}{c|}{\textbf{Training run with} $\mathbf{L_{Q-N+dedup}}$} \\
\cline{2-16}
 & \textbf{AuthPct$\uparrow$}  & $\mathbf{C_T}\uparrow$ & \textbf{FLD$\downarrow$} & \textbf{FID$\downarrow$} & \textbf{Q-N* $\downarrow$}  & \textbf{AuthPct$\uparrow$}  & $\mathbf{C_T}\uparrow$ & \textbf{FLD$\downarrow$} & \textbf{FID$\downarrow$} & \textbf{Q-N* $\downarrow$}  & \textbf{AuthPct$\uparrow$}  & $\mathbf{C_T}\uparrow$ & \textbf{FLD$\downarrow$} & \textbf{FID$\downarrow$} & \textbf{Q-N* $\downarrow$}  \\
\hline
2 & 30.65 & 11.75 & 6.26 & 624.22 & 3.721 & 30.09 & 12.19 & 6.43 & 640.68 & 3.730 & 30.76 & 11.55 & 7.13 & 692.01 & 1.643 \\
4 & 29.07 & 12.25 & 8.47 & 750.02 & 2.364 & 30.36 & 11.48 & 7.31 & 676.68 & 2.053 & 31.5 & 12.1 & 7.46 & 709.85 & 1.607 \\
6 & 30.82 & 11.99 & 7.92 & 738.6 & 2.043 & 30.04 & 12.32 & 8.07 & 723.56 & 1.840 & 30.63 & 12.58 & 8.08 & 741.31 & 1.228 \\
8 & 27.71 & 11.4 & 8.17 & 755.02 & 1.705 & 29.25 & 11.48 & 9.08 & 800.07 & 1.573 & 29.98 & 12.37 & 8.04 & 725.23 & 1.103 \\
\hline
\end{tabular}
}

\caption{Comparison of training framework performance evaluated on the entire CIFAR-10 dataset by averaging the scores obtained across the 10 latest model submissions. Column \# represents number of clients and * shows that the Q-N score is reported as x10\textsuperscript{-3}.}
\label{tab:cifar}
\end{table*}

In the second training setup, referred to as $L_{Q-N}$, the loss function used to evaluate and identify the top-performing model for further training is given by our quality-novelty score in equation \ref{ours}, where \textbf{$L_{Q-N}$} = Q-N score. 

The third setup, $L_{Q-N+dedup}$, builds upon the previous one by incorporating an additional step at the data loader level to exclude previously identified memorized training images from further model training. Training data duplication has been shown to increase memorization, therefore excluding samples that have previously been identified as memorized effectively creates a decentralized, privacy-preserving training data deduplication system. An image-level count of all identified instances of memorization is kept, which is then normalized via min-max normalization. Subsequently, the top 95th percentile of the most memorized samples are excluded from further training. On average, this results in the exclusion of 1.7\% of each node's local training set, aligning with findings in \cite{carlini2023extracting} and \cite{stein2023exposing}, which report a 2.5\%  rate of training data memorization at the end of model training. Our lower exclusion rate is attributed to the successful reduction of memorization and earlier intervention during training. 
%also our experiments train for less time than theirs%
This approach incurs no additional computational overhead, as memorized samples are already identified during model evaluation for further model training choice. To the best of our knowledge, we are the first to implement this method and demonstrate that excluding previously memorized samples effectively reduces training data memorization in subsequent training iterations.

All other training parameters remain identical across experiments, including duration and number of training epochs.
\subsection{Baseline memorization metrics}
We baseline using three recent metrics to measure memorization in GenAI models.

\textbf{AuthPct} \cite{authpct} deems each generated sample either authentic or inauthentic and returns the fraction of authentic samples. Inauthentic samples are those for which the distance to the nearest point in the training set is less than the distance between that training sample and its nearest neighbor in the training set.

$\mathbf{C_T}$ \cite{ctscore} analyzes the training, test, and generated sample sets and summarizes how often training samples are closer to generated samples than they are to test samples through a Mann-Whitney hypothesis test\cite{mann}. The data is partitioned into several cells using k-means clustering; the score is computed in each cell and averaged to obtain the final score.

\textbf{FLD} \cite{jiralerspong2023feature} is defined as the percentage of overfit Gaussians identified. The difference between the log-likelihoods under the training and test sets is calculated for each generated sample. The metric reports the percentage of generated samples for which the log-likelihood was higher under the training set than the test set. We
use the default implementation from \url{https://github.com/marcojira/fld}, which also incorporates the original implementations of AuthPct and $C_T$.

%A disadvantage of FLD and $C_T$ scores is that they utilize the test set in their calculation, relying on the assumption of non-overlapping training and test datasets, which is often not the case in practice. The CIFAR-10 dataset significantly overlaps the training and test set, with several images appearing in both sets. Since both FLD and $C_T$ scores compare the distribution of generated images to the training and test sets, the overlap can negatively impact their performance. This limitation underscores the need for further research to develop robust metrics for dataset overlap to provide accurate assessments of model memorization. Our proposed Q-N metric addresses this need by successfully identifying memorized samples without depending on the test set. 

\subsection{Evaluation compared to baseline scores}

\begin{figure*} [h!]
    \centering
    \includegraphics[width=\textwidth]{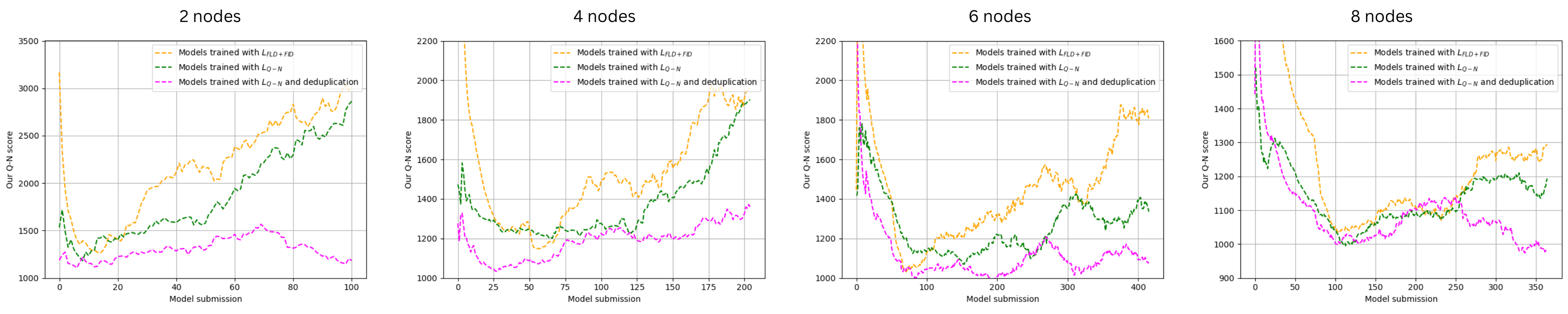}
    \caption{The lines represent the moving average of our proposed Q-N score, measured across the three training tasks - \textbf{Orange} used the loss function $L_{FLD+FID}$; \textbf{Green} trained with the objective $L_{Q-N}$; \textbf{Pink} trained with the objective $L_{Q-N}$ and excluded memorized training samples from further training at the dataloader level. We observe that training using our proposed loss function significantly reduces the extent of memorization in diffusion models, regardless of the number of participants. In addition, excluding memorized samples from further training significantly reduces memorization and is the most effective of the three methods.}
    \label{fig:blended_comparison}
\end{figure*}

In the first experiment, a pool of 100,000 generated images is created. The generated images are gradually replaced by training images until, finally, the training data makes up 100\% of the pool. Memorization between the training set and the image pool is measured using the Q-N, FLD, AuthPct, and $C_T$ scores. All scores trend in a direction indicating increased memorization. Our proposed Q-N score reaches an accuracy of 98.94\% for identifying memorized samples. In contrast, the other scores each operate on their own respective scales and provide measures of memorization, but they do not directly support the extraction of memorized sample pairs as our Q-N score does, thus preventing the calculation of an accuracy rate.

We run experiments training diffusion models according to the three previously outlined training setups:1) $L_{FLD+FID}$, 2) $L_{Q-N}$ and 3) $L_{Q-N+dedup}$. We calculate the mean of the Q-N score reported by each client node for each model submission and plot the results from the first and second training setups for comparison in Fig. \ref{fig:baseline}. The right shows the mean Q-N, FLD, AuthPct, and $C_T$ scores reported by each node for each model submission. Notably, while our proposed metric exhibits an upward trend, suggesting higher rates of memorization as model training progresses, the FLD, AuthPct, and $C_T$ scores tend to plateau or even trend in directions, indicating a decrease in memorization. Higher AuthPct and $C_T$ scores indicate less memorization, whereas lower FLD indicates less memorization. 

One explanation for the observed behavior of FLD, AuthPct, and $C_T$ scores is their sensitivity to the test set size. FLD and AuthPct require a minimum of 10,000 samples for an accurate evaluation and recommend using the entire train and test sets to compute the metric \cite{jiralerspong2023feature} — a method unsuitable for the federated learning scenario, where datasets are distributed across multiple nodes, and participants typically have varying amounts of data. Similarly, sample set size is a limitation for the $C_T$ score; without sufficient samples, the metric not only exhibits higher variance, but the ability to finely partition the instance space is reduced, which leads to mischaracterization and measurement errors \cite{ctscore}. In contrast, our proposed Q-N metric maintains accuracy regardless of sample set size, making it more cost-effective, robust and adaptable to real-world scenarios with limited resources, particularly within the federated learning context. Another limitation of the FLD and $C_T$ scores is that in addition to the training and generated sample set, they also utilize the test set for their calculation. This assumes that the training and test datasets are non-overlapping, which is often not the case in practice \cite{cifair}. Our proposed Q-N metric doesn't have this limitation, effectively measuring memorization without the need for the test set. 

The graphs in Fig. \ref{fig:baseline} show that our training method using $L_{Q-N}$ decreases memorization and is more sensitive to recognizing memorization behavior than all other metrics. 

On the left, all scores for the baseline experiment are shown and measured at various points in the training process. This model was trained traditionally, outside of a federated learning framework. As expected, memorization increases as training progresses. Unlike our proposed Q-N metric, the AuthPct, $C_T$, and FLD scores level out after only 1,000 epochs, failing to detect the memorization behavior the Q-N metric is able to identify as training progresses.

\subsection{Evaluating our objective function's effect on memorization}

Figure \ref{fig:blended_comparison} shows that for each experiment, training models using the proposed \textbf{$L_{Q-N}$} function based on our Q-N score succeeds in significantly reducing the extent of memorization in diffusion models, as compared to models trained using $L_{FLD+FID}$. In addition, the experiments show that excluding memorized samples from further training within the $L_{Q-N+dedup}$ training setup significantly reduces memorization and is the most effective of the three methods.

Table \ref{tab:cifar} presents the mean scores of the 10 latest model submissions across all training setups, evaluated using the entire CIFAR-10 dataset rather than individual node data splits. We draw three conclusions: 1) that training using our proposed methods is successful in decreasing the extent of memorization while all other metrics, including FID, remain roughly the same, 2) that the more nodes there are and the fewer data they each hold, the longer it takes for a model to reach the same image generation quality. However, the level of memorization exhibited is significantly lower. 3) the level of memorization decreases the more participants are involved in training, even when our proposed objective function isn't used in training but our proposed federated learning framework is.

\section{Conclusion}
Our experiments demonstrate that the proposed decentralized, asynchronous federated learning framework (here experimentally implemented using a blockchain), in conjunction with the Q-N score, successfully reduces training data memorization in diffusion models. We leverage the decentralized, leaderless nature of public blockchains, enabling participants to operate asynchronously and independently with local training strategies, while also considering validation-only participants to foster broader participation. Our proposed metric shows greater sensitivity in detecting memorization behavior across various training setups, outperforming concurrent metrics. These findings underscore the importance of incorporating novel evaluation strategies, like our proposed Q-N score, to ensure training data privacy and enhance the generalization of GenAI models. Future efforts should focus on integrating our loss function within the training code of diffusion models and further exploring decentralised federated learning solutions to enhance privacy, transparency, and efficiency for ethical and collaborative AI training. PDFed has the potential to reshape the creative industry by fostering a more equitable AI ecosystem for artists and creators. Thus, future work should study the socio-economic drivers necessary to ensure the adoption and sustainability of truly decentralised, peer-to-peer GenAI training, especially considering input from the creative sector.

\begin{acks}
PDFed was supported by DECaDE under EPSRC Grant EP/T022485/1.
\end{acks}
\bibliographystyle{ACM-Reference-Format}
\bibliography{software}

%%% -*-BibTeX-*-
%%% Do NOT edit. File created by BibTeX with style
%%% ACM-Reference-Format-Journals [18-Jan-2012].

\begin{thebibliography}{28}

%%% ====================================================================
%%% NOTE TO THE USER: you can override these defaults by providing
%%% customized versions of any of these macros before the \bibliography
%%% command.  Each of them MUST provide its own final punctuation,
%%% except for \shownote{}, \showDOI{}, and \showURL{}.  The latter two
%%% do not use final punctuation, in order to avoid confusing it with
%%% the Web address.
%%%
%%% To suppress output of a particular field, define its macro to expand
%%% to an empty string, or better, \unskip, like this:
%%%
%%% \newcommand{\showDOI}[1]{\unskip}   % LaTeX syntax
%%%
%%% \def \showDOI #1{\unskip}           % plain TeX syntax
%%%
%%% ====================================================================

\ifx \showCODEN    \undefined \def \showCODEN     #1{\unskip}     \fi
\ifx \showDOI      \undefined \def \showDOI       #1{#1}\fi
\ifx \showISBNx    \undefined \def \showISBNx     #1{\unskip}     \fi
\ifx \showISBNxiii \undefined \def \showISBNxiii  #1{\unskip}     \fi
\ifx \showISSN     \undefined \def \showISSN      #1{\unskip}     \fi
\ifx \showLCCN     \undefined \def \showLCCN      #1{\unskip}     \fi
\ifx \shownote     \undefined \def \shownote      #1{#1}          \fi
\ifx \showarticletitle \undefined \def \showarticletitle #1{#1}   \fi
\ifx \showURL      \undefined \def \showURL       {\relax}        \fi
% The following commands are used for tagged output and should be
% invisible to TeX
\providecommand\bibfield[2]{#2}
\providecommand\bibinfo[2]{#2}
\providecommand\natexlab[1]{#1}
\providecommand\showeprint[2][]{arXiv:#2}

\bibitem[Alaa et~al\mbox{.}(2021)]%
        {authpct}
\bibfield{author}{\bibinfo{person}{Ahmed~M. Alaa}, \bibinfo{person}{Boris van Breugel}, \bibinfo{person}{Evgeny~S. Saveliev}, {and} \bibinfo{person}{Mihaela van~der Schaar}.} \bibinfo{year}{2021}\natexlab{}.
\newblock \showarticletitle{How Faithful is your Synthetic Data? Sample-level Metrics for Evaluating and Auditing Generative Models}. In \bibinfo{booktitle}{\emph{International Conference on Machine Learning}}.
\newblock


\bibitem[Balan et~al\mbox{.}(2023a)]%
        {balan2023ekila}
\bibfield{author}{\bibinfo{person}{Kar Balan}, \bibinfo{person}{Shruti Agarwal}, \bibinfo{person}{Simon Jenni}, \bibinfo{person}{Andy Parsons}, \bibinfo{person}{Andrew Gilbert}, {and} \bibinfo{person}{John Collomosse}.} \bibinfo{year}{2023}\natexlab{a}.
\newblock \showarticletitle{EKILA: Synthetic Media Provenance and Attribution for Generative Art}. In \bibinfo{booktitle}{\emph{2023 IEEE/CVF Conference on Computer Vision and Pattern Recognition Workshops (CVPRW)}}. \bibinfo{publisher}{IEEE Computer Society}, \bibinfo{address}{Los Alamitos, CA, USA}, \bibinfo{pages}{913--922}.
\newblock


\bibitem[Balan et~al\mbox{.}(2023b)]%
        {decorait}
\bibfield{author}{\bibinfo{person}{Kar Balan}, \bibinfo{person}{Andrew Gilbert}, \bibinfo{person}{Alexander Black}, \bibinfo{person}{Simon Jenni}, \bibinfo{person}{Andy Parsons}, {and} \bibinfo{person}{John Collomosse}.} \bibinfo{year}{2023}\natexlab{b}.
\newblock \showarticletitle{DECORAIT - DECentralized Opt-in/out Registry for AI Training}. In \bibinfo{booktitle}{\emph{Proceedings of the 20th ACM SIGGRAPH European Conference on Visual Media Production}} \emph{(\bibinfo{series}{CVMP '23})}. \bibinfo{publisher}{Association for Computing Machinery}, \bibinfo{address}{New York, NY, USA}, Article \bibinfo{articleno}{4}, \bibinfo{numpages}{10}~pages.
\newblock
\showISBNx{9798400704260}


\bibitem[Barz and Denzler(2020)]%
        {cifair}
\bibfield{author}{\bibinfo{person}{Björn Barz} {and} \bibinfo{person}{Joachim Denzler}.} \bibinfo{year}{2020}\natexlab{}.
\newblock \showarticletitle{Do We Train on Test Data? Purging CIFAR of Near-Duplicates}.
\newblock \bibinfo{journal}{\emph{Journal of Imaging}} \bibinfo{volume}{6}, \bibinfo{number}{6} (\bibinfo{date}{June} \bibinfo{year}{2020}), \bibinfo{pages}{41}.
\newblock


\bibitem[Behera et~al\mbox{.}(2022)]%
        {behera2022federated}
\bibfield{author}{\bibinfo{person}{Monik~Raj Behera}, \bibinfo{person}{Sudhir Upadhyay}, {and} \bibinfo{person}{Suresh Shetty}.} \bibinfo{year}{2022}\natexlab{}.
\newblock \bibinfo{title}{Federated Learning using Smart Contracts on Blockchains, based on Reward Driven Approach}.
\newblock
\newblock
\showeprint[arxiv]{2107.10243}~[cs.CR]


\bibitem[Black et~al\mbox{.}(2021)]%
        {Black_2021_CVPR}
\bibfield{author}{\bibinfo{person}{Alexander Black}, \bibinfo{person}{Tu Bui}, \bibinfo{person}{Hailin Jin}, \bibinfo{person}{Vishy Swaminathan}, {and} \bibinfo{person}{John Collomosse}.} \bibinfo{year}{2021}\natexlab{}.
\newblock \showarticletitle{Deep Image Comparator: Learning To Visualize Editorial Change}. In \bibinfo{booktitle}{\emph{Proceedings of the IEEE/CVF Conference on Computer Vision and Pattern Recognition (CVPR) Workshops}}. \bibinfo{pages}{972--980}.
\newblock


\bibitem[Bui et~al\mbox{.}(2019)]%
        {archangel}
\bibfield{author}{\bibinfo{person}{Tu Bui}, \bibinfo{person}{Daniel Cooper}, \bibinfo{person}{John Collomosse}, \bibinfo{person}{Mark Bell}, \bibinfo{person}{Alex Green}, \bibinfo{person}{John Sheridan}, \bibinfo{person}{Jez Higgins}, \bibinfo{person}{Arindra Das}, \bibinfo{person}{Jared Keller}, \bibinfo{person}{Olivier Thereaux}, {and} \bibinfo{person}{Alan Brown}.} \bibinfo{year}{2019}\natexlab{}.
\newblock \showarticletitle{ARCHANGEL: Tamper-Proofing Video Archives Using Temporal Content Hashes on the Blockchain}. In \bibinfo{booktitle}{\emph{2019 IEEE/CVF Conference on Computer Vision and Pattern Recognition Workshops (CVPRW)}}. \bibinfo{publisher}{IEEE Computer Society}, \bibinfo{address}{Los Alamitos, CA, USA}, \bibinfo{pages}{2793--2801}.
\newblock


\bibitem[Carlini et~al\mbox{.}(2023)]%
        {carlini2023extracting}
\bibfield{author}{\bibinfo{person}{Nicholas Carlini}, \bibinfo{person}{Jamie Hayes}, \bibinfo{person}{Milad Nasr}, \bibinfo{person}{Matthew Jagielski}, \bibinfo{person}{Vikash Sehwag}, \bibinfo{person}{Florian Tram\`{e}r}, \bibinfo{person}{Borja Balle}, \bibinfo{person}{Daphne Ippolito}, {and} \bibinfo{person}{Eric Wallace}.} \bibinfo{year}{2023}\natexlab{}.
\newblock \showarticletitle{Extracting training data from diffusion models}. In \bibinfo{booktitle}{\emph{Proceedings of the 32nd USENIX Conference on Security Symposium}} (Anaheim, CA, USA) \emph{(\bibinfo{series}{SEC '23})}. \bibinfo{publisher}{USENIX Association}, \bibinfo{address}{USA}, Article \bibinfo{articleno}{294}, \bibinfo{numpages}{18}~pages.
\newblock
\showISBNx{978-1-939133-37-3}


\bibitem[Chen et~al\mbox{.}(2020)]%
        {chen2020simple}
\bibfield{author}{\bibinfo{person}{Ting Chen}, \bibinfo{person}{Simon Kornblith}, \bibinfo{person}{Mohammad Norouzi}, {and} \bibinfo{person}{Geoffrey Hinton}.} \bibinfo{year}{2020}\natexlab{}.
\newblock \showarticletitle{A simple framework for contrastive learning of visual representations}. In \bibinfo{booktitle}{\emph{International conference on machine learning}}. PMLR, \bibinfo{pages}{1597--1607}.
\newblock


\bibitem[{Coalition for Content Provenance and Authenticity}(2021)]%
        {c2pa}
\bibfield{author}{\bibinfo{person}{{Coalition for Content Provenance and Authenticity}}.} \bibinfo{year}{2021}\natexlab{}.
\newblock \bibinfo{booktitle}{\emph{Draft Technical Specification 0.7}}.
\newblock \bibinfo{type}{{T}echnical {R}eport}. \bibinfo{institution}{C2PA}.
\newblock
\urldef\tempurl%
\url{https://c2pa.org/public-draft/}
\showURL{%
\tempurl}


\bibitem[Hendrycks and Dimanetterich(2019)]%
        {hendrycks2019robustness}
\bibfield{author}{\bibinfo{person}{Dan Hendrycks} {and} \bibinfo{person}{Thomas Dimanetterich}.} \bibinfo{year}{2019}\natexlab{}.
\newblock \showarticletitle{Benchmarking Neural Network Robustness to Common Corruptions and Perturbations}.
\newblock \bibinfo{journal}{\emph{Proceedings of the International Conference on Learning Representations}} (\bibinfo{year}{2019}).
\newblock


\bibitem[Ho et~al\mbox{.}(2020)]%
        {hddpm}
\bibfield{author}{\bibinfo{person}{Jonathan Ho}, \bibinfo{person}{Ajay Jain}, {and} \bibinfo{person}{Pieter Abbeel}.} \bibinfo{year}{2020}\natexlab{}.
\newblock \bibinfo{title}{Denoising Diffusion Probabilistic Models}.
\newblock
\newblock
\showeprint[arxiv]{2006.11239}~[cs.LG]


\bibitem[Jiralerspong et~al\mbox{.}(2023)]%
        {jiralerspong2023feature}
\bibfield{author}{\bibinfo{person}{Marco Jiralerspong}, \bibinfo{person}{Joey Bose}, \bibinfo{person}{Ian Gemp}, \bibinfo{person}{Chongli Qin}, \bibinfo{person}{Yoram Bachrach}, {and} \bibinfo{person}{Gauthier Gidel}.} \bibinfo{year}{2023}\natexlab{}.
\newblock \showarticletitle{Feature Likelihood Score: Evaluating the Generalization of Generative Models Using Samples}. In \bibinfo{booktitle}{\emph{Thirty-seventh Conference on Neural Information Processing Systems}}.
\newblock


\bibitem[Krizhevsky(2009)]%
        {cifar10}
\bibfield{author}{\bibinfo{person}{Alex Krizhevsky}.} \bibinfo{year}{2009}\natexlab{}.
\newblock \bibinfo{booktitle}{\emph{Learning multiple layers of features from tiny images}}.
\newblock \bibinfo{type}{{T}echnical {R}eport}.
\newblock


\bibitem[Mann and Whitney(1947)]%
        {mann}
\bibfield{author}{\bibinfo{person}{Henry~Berthold Mann} {and} \bibinfo{person}{Donald~Ransom Whitney}.} \bibinfo{year}{1947}\natexlab{}.
\newblock \showarticletitle{{On a Test of Whether one of Two Random Variables is Stochastically Larger than the Other}}.
\newblock \bibinfo{journal}{\emph{The Annals of Mathematical Statistics}} \bibinfo{volume}{18}, \bibinfo{number}{1} (\bibinfo{year}{1947}), \bibinfo{pages}{50 -- 60}.
\newblock
\urldef\tempurl%
\url{https://doi.org/10.1214/aoms/1177730491}
\showDOI{\tempurl}


\bibitem[Meehan et~al\mbox{.}(2020)]%
        {ctscore}
\bibfield{author}{\bibinfo{person}{Casey Meehan}, \bibinfo{person}{Kamalika Chaudhuri}, {and} \bibinfo{person}{Sanjoy Dasgupta}.} \bibinfo{year}{2020}\natexlab{}.
\newblock \showarticletitle{A non-parametric test to detect data-copying in generative models}.
\newblock \bibinfo{journal}{\emph{International Conference on Artificial Intelligence and Statistics}} (\bibinfo{date}{Apr} \bibinfo{year}{2020}).
\newblock


\bibitem[Mugunthan et~al\mbox{.}(2020)]%
        {blockflow}
\bibfield{author}{\bibinfo{person}{Vaikkunth Mugunthan}, \bibinfo{person}{Ravi Rahman}, {and} \bibinfo{person}{Lalana Kagal}.} \bibinfo{year}{2020}\natexlab{}.
\newblock \bibinfo{title}{BlockFLow: An Accountable and Privacy-Preserving Solution for Federated Learning}.
\newblock
\newblock
\showeprint[arxiv]{2007.03856}~[cs.LG]


\bibitem[Oquab et~al\mbox{.}(2024)]%
        {dinov2}
\bibfield{author}{\bibinfo{person}{Maxime Oquab}, \bibinfo{person}{Timoth{\'e}e Darcet}, \bibinfo{person}{Th{\'e}o Moutakanni}, \bibinfo{person}{Huy~V. Vo}, \bibinfo{person}{Marc Szafraniec}, \bibinfo{person}{Vasil Khalidov}, \bibinfo{person}{Pierre Fernandez}, \bibinfo{person}{Daniel HAZIZA}, \bibinfo{person}{Francisco Massa}, \bibinfo{person}{Alaaeldin El-Nouby}, \bibinfo{person}{Mido Assran}, \bibinfo{person}{Nicolas Ballas}, \bibinfo{person}{Wojciech Galuba}, \bibinfo{person}{Russell Howes}, \bibinfo{person}{Po-Yao Huang}, \bibinfo{person}{Shang-Wen Li}, \bibinfo{person}{Ishan Misra}, \bibinfo{person}{Michael Rabbat}, \bibinfo{person}{Vasu Sharma}, \bibinfo{person}{Gabriel Synnaeve}, \bibinfo{person}{Hu Xu}, \bibinfo{person}{Herve Jegou}, \bibinfo{person}{Julien Mairal}, \bibinfo{person}{Patrick Labatut}, \bibinfo{person}{Armand Joulin}, {and} \bibinfo{person}{Piotr Bojanowski}.} \bibinfo{year}{2024}\natexlab{}.
\newblock \showarticletitle{{DINO}v2: Learning Robust Visual Features without Supervision}.
\newblock \bibinfo{journal}{\emph{Transactions on Machine Learning Research}} (\bibinfo{year}{2024}).
\newblock
\showISSN{2835-8856}


\bibitem[Ramanan and Nakayama(2020)]%
        {baffle}
\bibfield{author}{\bibinfo{person}{Paritosh Ramanan} {and} \bibinfo{person}{Kiyoshi Nakayama}.} \bibinfo{year}{2020}\natexlab{}.
\newblock \showarticletitle{BAFFLE : Blockchain Based Aggregator Free Federated Learning}. In \bibinfo{booktitle}{\emph{2020 IEEE International Conference on Blockchain (Blockchain)}}. \bibinfo{publisher}{IEEE Computer Society}, \bibinfo{address}{Los Alamitos, CA, USA}, \bibinfo{pages}{72--81}.
\newblock


\bibitem[Somepalli et~al\mbox{.}(2023)]%
        {somepalli2022diffusion}
\bibfield{author}{\bibinfo{person}{Gowthami Somepalli}, \bibinfo{person}{Vasu Singla}, \bibinfo{person}{Micah Goldblum}, \bibinfo{person}{Jonas Geiping}, {and} \bibinfo{person}{Tom Goldstein}.} \bibinfo{year}{2023}\natexlab{}.
\newblock \showarticletitle{Diffusion Art or Digital Forgery? Investigating Data Replication in Diffusion Models}. In \bibinfo{booktitle}{\emph{2023 IEEE/CVF Conference on Computer Vision and Pattern Recognition (CVPR)}}. \bibinfo{publisher}{IEEE Computer Society}, \bibinfo{address}{Los Alamitos, CA, USA}, \bibinfo{pages}{6048--6058}.
\newblock


\bibitem[Stein et~al\mbox{.}(2023)]%
        {stein2023exposing}
\bibfield{author}{\bibinfo{person}{George Stein}, \bibinfo{person}{Jesse~C. Cresswell}, \bibinfo{person}{Rasa Hosseinzadeh}, \bibinfo{person}{Yi Sui}, \bibinfo{person}{Brendan~Leigh Ross}, \bibinfo{person}{Valentin Villecroze}, \bibinfo{person}{Zhaoyan Liu}, \bibinfo{person}{Anthony~L. Caterini}, \bibinfo{person}{Eric Taylor}, {and} \bibinfo{person}{Gabriel Loaiza-Ganem}.} \bibinfo{year}{2023}\natexlab{}.
\newblock \showarticletitle{Exposing flaws of generative model evaluation metrics and their unfair treatment of diffusion models}. In \bibinfo{booktitle}{\emph{Thirty-seventh Conference on Neural Information Processing Systems}}.
\newblock


\bibitem[Tolias et~al\mbox{.}(2016)]%
        {tolias2015particular}
\bibfield{author}{\bibinfo{person}{Giorgos Tolias}, \bibinfo{person}{Ronan Sicre}, {and} \bibinfo{person}{Herv{\'{e}} J{\'{e}}gou}.} \bibinfo{year}{2016}\natexlab{}.
\newblock \showarticletitle{Particular object retrieval with integral max-pooling of {CNN} activations}. In \bibinfo{booktitle}{\emph{4th International Conference on Learning Representations, {ICLR} 2016, San Juan, Puerto Rico, May 2-4, 2016, Conference Track Proceedings}}.
\newblock


\bibitem[Tomiyama et~al\mbox{.}(2023)]%
        {competitive}
\bibfield{author}{\bibinfo{person}{Eisuke Tomiyama}, \bibinfo{person}{Hiroshi Esaki}, {and} \bibinfo{person}{Hideya Ochiai}.} \bibinfo{year}{2023}\natexlab{}.
\newblock \showarticletitle{Competitive and Asynchronous Decentralized Federated Learning with Blockchain Smart Contracts}. In \bibinfo{booktitle}{\emph{Proceedings of the 2023 ACM Conference on Information Technology for Social Good}} \emph{(\bibinfo{series}{GoodIT '23})}. \bibinfo{pages}{92–99}.
\newblock


\bibitem[Tun et~al\mbox{.}(2023)]%
        {sensitivevis}
\bibfield{author}{\bibinfo{person}{Ye~Lin Tun}, \bibinfo{person}{Chu~Myaet Thwal}, \bibinfo{person}{Ji~Su Yoon}, \bibinfo{person}{Sun~Moo Kang}, \bibinfo{person}{Chaoning Zhang}, {and} \bibinfo{person}{Choong~Seon Hong}.} \bibinfo{year}{2023}\natexlab{}.
\newblock \bibinfo{title}{Federated Learning with Diffusion Models for Privacy-Sensitive Vision Tasks}.
\newblock
\newblock
\showeprint[arxiv]{2311.16538}~[cs.LG]


\bibitem[Yang et~al\mbox{.}(2023)]%
        {feddisc}
\bibfield{author}{\bibinfo{person}{Mingzhao Yang}, \bibinfo{person}{Shangchao Su}, \bibinfo{person}{Bin Li}, {and} \bibinfo{person}{Xiangyang Xue}.} \bibinfo{year}{2023}\natexlab{}.
\newblock \bibinfo{title}{Exploring One-shot Semi-supervised Federated Learning with A Pre-trained Diffusion Model}.
\newblock
\newblock
\showeprint[arxiv]{2305.04063}~[cs.CV]


\bibitem[Yoon et~al\mbox{.}(2023)]%
        {yoon2023diffusion}
\bibfield{author}{\bibinfo{person}{TaeHo Yoon}, \bibinfo{person}{Joo~Young Choi}, \bibinfo{person}{Sehyun Kwon}, {and} \bibinfo{person}{Ernest~K. Ryu}.} \bibinfo{year}{2023}\natexlab{}.
\newblock \showarticletitle{Diffusion Probabilistic Models Generalize when They Fail to Memorize}. In \bibinfo{booktitle}{\emph{ICML 2023 Workshop on Structured Probabilistic Inference {\&} Generative Modeling}}.
\newblock
\urldef\tempurl%
\url{https://openreview.net/forum?id=shciCbSk9h}
\showURL{%
\tempurl}


\bibitem[Yun et~al\mbox{.}(2023)]%
        {bcafl}
\bibfield{author}{\bibinfo{person}{Jian Yun}, \bibinfo{person}{Yusheng Lu}, {and} \bibinfo{person}{Xinyu Liu}.} \bibinfo{year}{2023}\natexlab{}.
\newblock \showarticletitle{BCAFL: A Blockchain-Based Framework for Asynchronous Federated Learning Protection}.
\newblock \bibinfo{journal}{\emph{Electronics}} \bibinfo{volume}{12}, \bibinfo{number}{20} (\bibinfo{year}{2023}).
\newblock
\showISSN{2079-9292}
\urldef\tempurl%
\url{https://www.mdpi.com/2079-9292/12/20/4214}
\showURL{%
\tempurl}


\bibitem[Zhao et~al\mbox{.}(2024)]%
        {noniid}
\bibfield{author}{\bibinfo{person}{Zhuang Zhao}, \bibinfo{person}{Feng Yang}, {and} \bibinfo{person}{Guirong Liang}.} \bibinfo{year}{2024}\natexlab{}.
\newblock \showarticletitle{Federated Learning Based on Diffusion Model to Cope with Non-IID Data}. In \bibinfo{booktitle}{\emph{Pattern Recognition and Computer Vision}}. \bibinfo{publisher}{Springer Nature Singapore}, \bibinfo{address}{Singapore}, \bibinfo{pages}{220--231}.
\newblock
\showISBNx{978-981-99-8546-3}


\end{thebibliography}

\end{document}